\documentclass[11pt]{article}
\usepackage[margin=3cm]{geometry}
\usepackage{amsmath, amsthm}
\usepackage{natbib}
\usepackage{graphicx}
\usepackage{epstopdf}
\usepackage{epsfig}
\usepackage[english]{babel}
\usepackage[all]{xy}
\usepackage{natbib}

\newcommand{\tr}{^{\prime}}
\def\b#1{\mbox{\boldmath $#1$}}    
\def\bl#1{\mbox{\footnotesize \boldmath {$#1$}}} 

\newcommand{\diag}{{\rm diag}}    
\title{Fitting Directed Acyclic Graphs with latent nodes as finite mixtures models, with application to education transmission}
\author{Antonio Forcina, Dip. Economia, Finanza e Statistica, Univ. of Perugia, Italy \\
        Salvatore Modica,  Dip. di Scienze Economiche, Aziendali e Finanziarie, Univ. of Palermo, Italy}
\begin{document}
\maketitle
\begin{abstract}
This paper describes an efficient EM algorithm for maximum likelihood estimation of a system of non linear structural equations corresponding to a directed acyclic graph model; this can contain an arbitrary number of latent variables, as long as the model is identifiable. The only limitation is that the endogenous variables in the model must be discrete, with qualitative or ordered categories, while the exogenous variables may be arbitrary. The models described in this paper are defined by the list of structural equations and, for each equation, the type of link function and the linear model to be used. These models are an extended version of finite mixture models which may be suitable for causal inference when several sources of latent heterogeneity may be present. An application to the problem of education transmission, where one would like to control for the ability of the parents and that of the child, is presented as an illustration.
\end{abstract}
\section{Introduction}
\label{intro}
Structural equation models (SEM) are defined by a system of nonlinear equations specifying which variables have a direct causal effect on each endogenous variable in the system. A recursive non parametric SEM is equivalent to a directed acyclic graph (DAG) and, also, to a set of conditional independence statements. \cite{Pearl95}, among others, has shown that, under certain conditions (the {\em back-door} and the {\em front-door} criteria), causal effects can be estimated from the frequency distribution of the observed variables; these conditions are, however, rather restrictive and are difficult to combine with statistical modeling assumptions. In this paper we restrict attention to situations where the joint distribution of the observed variables is identifiable because of suitable modeling restrictions and we describe an efficient algorithm for maximum likelihood estimation. Certain routines of this  algorithm may also be used to compute one of the most common measure of direct causal effects known as {\em natural} direct effect \cite{PearlNDE}.

The class of models considered in this paper may be seen as an extension of latent class models in the sense that observable variables need not be independent conditionally on the latent ones. In addition, an observable variable may have a direct effect on a latent one and a latent variable may have a direct effect on an other latent which is conceptually distinct. These models are not entirely new, for example \cite{Hag:2002} has considered an application to a social science context of a model which is a special case of those considered here. The class of mixture models considered by \cite{AlfTrov} may be seen as a special case of those studied here, relative to the dependence structure; for a more detailed discussion of this relationship, see section \ref{Disc}.

We present an application in the context of education transmission, a much debated issue in Econometrics and Labor Economics. In order to assess the causal effect of the education of the parents on that of their child, one needs to control for the latent endowments of the parents and that of the child, which are likely to be strongly associated. The approach we propose is based on estimating a recursive system of structural equations where the natural endowment of parents and child are treated as two latent endogenous variables; this, we believe, provides an innovative contribution to the existing literature on the subject which we review briefly in Section 5.

The class of models studied in this paper are defined in section 2 where we examine the relationship with related models. The computation of maximum likelihood estimates and their implementation are discussed in section 3, an approach to the evaluation of causal effects is presented in section 4 and the application to education transmission is presented in section 5.
\section{A class of semi-parametric structural equation models}
\label{sec:model}
We recall, following \cite{PearlB}, that a non parametric recursive structural equation model is a system of equations in the variables $Z_1,\dots ,Z_n$
\begin{equation}
\label{eq:sem} Z_i=f_i(pa_i,\epsilon_i), \quad i=1,\dots ,n \end{equation}
where $pa_i$ is the subset of variables which are assumed to be the direct causes of $Z_i$, these are usually called {\it parents}, and $\epsilon_1,\dots ,\epsilon_n$ is a set of independent background exogenous variables which account for all residual effects. The fact that the system is recursive implies that, if $Z_h$ is a parent of $Z_i$, then $h<i$; in other words, $Z_h$ cannot have a direct effect on $Z_i$ if $i<h$. The system is non-parametric in the sense that the distribution of the $\epsilon_i$s and the form of the functions $f_i$ do not need to be specified. Such a system of equations is equivalent to a directed acyclic graph (shortened to DAG in the following) where endogenous variables are represented by nodes and there is an arrow from $Z_h$ to $Z_i$ if $Z_h$ is a direct cause of $Z_i$, that is if $Z_h\in pa_i$. A convenient property of causal DAGs is that the joint distribution may be factorized into the product of the conditional distribution of each node given its parents. A DAG can contain one or more latent nodes, for example in the case of education transmission to be discussed in section \ref{sec:app}, the unobservable endowments of the parents and that of the child are supposed to affect the educational achievements of the latter.

The methodology described in this paper exploits recent results on marginal models for discrete data, thus it is applicable when endogenous variables, observed or latent, are categorical or have been turned into discrete. Our models differ from non parametric SEM because, when a variable is assumed to depend on two or more other variables, we allow some of these effects to be additive on a logit scale appropriate to the nature of the response variable under consideration. Essentially, logits of type {\it reference category, (r)} or {\it adjacent, (a)} are more appropriate when response categories are not ordered, logits of type {\it global, (g)} are preferable when response categories are ordered and logits of type  {\it continuation, (c)} are more suitable when response categories correspond to survival or achievements, see \cite{ColFor:01} for a detailed discussion. If $Z_i$ has categories coded as $0,1,\dots ,c_i-1$, the $i$th structural equation has $c_i-1$ components, one for each logit of  $Z_i$ and, in the special case when the effects of its parents are additive, the $h$th logit ($h=1,\ldots , c_i-1$) may be written as
\begin{equation}
\lambda_{ih} = \sum_{l=1}^{h} \beta_{i0l}+\sum_{Z_j\in pa_i} \sum_{l=1}^{c_j-1} \beta_{ijl}\: I(Z_j\geq l),
\label{eq:struct}
\end{equation}
where $I(Z_j\geq l)$ is the indicator function. In the equation above we have used the incremental coding for the $\beta$s, this means that, for instance, $\beta_{i0h}$ is the difference in the intercepts of the $h$ and $h-1$ logits for $Z_i$ while $\beta_{ijl}$ measures the increase in each logit $\lambda_{ih}$ when $Z_j$ is changed from $l-1$ to $l$. The reconstruction formulas for the case of logits global (g) and adjacent (a), the only types used in this paper, are given below
\begin{eqnarray*}
(g):\: P(Z_i=h)&=&\frac{\exp(\lambda_{ih})}{1+\exp(\lambda_{ih})}-
\frac{\exp(\lambda_{i,h-1})}{1+\exp(\lambda_{i,h-1})} \\
(a):\: P(Z_i=h) &=& \frac{\exp(\sum_{l=1}^h\lambda_{il})} {1+\sum_{h=1}^{c_i-1}\exp(\sum_{l=1}^h\lambda_{il})}
\end{eqnarray*}

In the user-interface we have designed, any model of our class can be specified by the following data:
\begin{itemize}
\item An ordered list of the endogenous variables such that, if there is an arrow from $Z_i$ to $Z_j$, then $i< j$; this must be consistent with the lexicographic order of the observed frequencies,meaning that the categories of $Z_j$ run faster of those of $Z_i$ ;
\item A binary indicator $b_i$ which is set to 1 if $Z_i$ is latent;
\item For each endogenous variable, the list of its parents; the collection of these lists determines the DAG;
\item For each node, the corresponding link function; this is determined by the number of categories of the node variable and the type of logit (adjacent, global, continuation) which determines how its conditional distribution is parameterized;
\item For each endogenous variable, a regression model which specifies how its logits depend on the parents and, possibly, on additional exogenous variables measured at the level of statistical units; this is determined by a design matrix for each response variable.
\end{itemize}
\subsection{Identifiability}
\label{Identi}
Identifiability results for latent class models under conditional independence are by now well established.
Recent results by \cite{Allman} can handle several extended latent class models where certain subsets of the observable variables may be associated conditionally to the latent. However, to our knowledge, no results are available to determine whether a general DAG with an arbitrary number of latent variables is identifiable under a specified set of modeling restrictions. For this reason, in the application we have used the numerical method described by \cite{Forc08} which can detect, with probability arbitrarily close to 1, whether a given model is locally identifiable everywhere in the parameter space.
Essentially, the method consists in sampling a sufficiently large number of points from the parameter space and computes the jacobian matrix obtained by differentiating the log-linear parameters of the saturated model for the joint distribution of the observable variables with respect to the actual parameters of the model. If, in a reasonably large sample, the inverse condition number never goes below, say, $10^{-10}$, we can be highly confident that the model is locally identified almost everywhere in the parameter space.

Typical modeling restrictions that might be used to achieve identifiability are assumptions of additivity within a given link function, like, for example, a multivariate logistic function. Continuous covariates may be included as exogenous variables; these are the variables determined outside the system so that there is no equation that describes their behavior. Clearly, when continuous covariates are available, a linear regression model within the assumed link function must be used.
\subsection{Discussion}
\label{Disc}
An interesting instance of the models described above was used informally by \cite{Hag:2002} as an extended latent class model. It may be interesting to note that, while in a basic latent class model, where the latent has no parents, the parameters which determine the marginal distribution of the latent are somehow separate from those which determine the conditional distribution of the responses, in the general context described here, in principle, any node of the DAG may correspond to a latent variable and, if there is a latent node $Z_i$ which has no parents, its marginal distribution is determined by the $\beta_{i0h}$, the intercept parameters for the adjacent logits, whose number equals the number of latent categories minus 1.

A different, but closely related literature is that based on finite mixture models, like, for instance, the models studied in \cite{AlfTrov} where a selection variable and two or more response variables are assumed to depend on a multivariate continuous latent distribution. However, when the latent distribution is approximated with a discrete distribution with $K$ support points, the resulting model is equivalent to a DAG model with a single discrete latent variable, say $U$, having $K$ categories; the special case where there are two responses $Y_1,\:Y_2$ and a selection variable $Y_0$ is displayed in the DAG below
$$
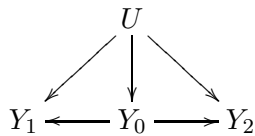
\xymatrix{
 & U \ar@{->}[dl] \ar@{->}[dr] \ar@{->}[d] &  \\
Y_1 & Y_0 \ar@{->}[l] \ar@{->}[r] & Y_2
}
$$
It is worth noting that the true multivariate nature of the underlying latent, once turned into a discrete one, should show up in the values of the estimated intercept parameters $\beta_{iul}$, where $i$ indexes the response variable, $u$ the single latent and $l$ the category of the latent. For instance, the fact that $\beta_{iul}$ is positive (or negative) for all $l$ indicates that the underlying latent is essentially uni-dimensional and that its categories may be ordered from best to worst in a unique way.
\section{Maximum likelihood estimation}
\label{sec:MLe}
In principle, under the assumption that, conditionally on exogenous covariates, the joint latent distribution of the variables in the DAG is multinomial, any identifiable model may be fitted by an EM algorithm. In the E-step, one reconstructs the hypothetical frequency table of the full joint distribution on the basis of the posterior probabilities; then, in the M-step, one maximizes the multinomial likelihood of the full latent distribution specified by the DAG.

More formally, in the E-step: we may treat the collection of latent variables in the DAG as if it were a single variable with a number of categories equal to the number of cells in their joint distribution. Let $\b u$ specify a cell within the joint distribution of the latent variables and $\b m$ a cell within the joint distribution of the manifest variables; let $\hat\pi_{\bl u,\bl m}(i)$ denote the probability of belonging to latent configuration $\b u$ and having observed configuration $\b m$ for the $i$th unit, where the dependence on $i$ can be omitted if there are no individual level covariates. Let $N_{\bl m}$ denotes the observed frequency in cell $\b m$; the entries of the reconstructed frequency table are given by
$$
\psi_{\bl m,\bl u} = N_{\bl m}\,\frac{\hat\pi_{\bl u ,\bl m}}{\sum_{\bl u} \hat\pi_{\bl u ,\bl m}}.
$$
However, because the elements of $\hat\pi_{\bl u ,\bl m}$ are arranged in the lexicographic order determined by the DAG and latent variables may be in any position, an efficient way of implementing the EM algorithm within a general DAG which can contain an arbitrary number of latent variables may be based on matrix notations and rearrangement indices. Suppose that probabilities are stored as elements of the vector $\b \pi$ in the lexicographic order defined by the DAG and let $\b j_t$ be the vector of indices which transforms $\b \pi$ into $\b \Pi$, the two way table with the categories of the latent arranged in lexicographic order by row and those of the observed variables by column. Then the E-step can be written in matrix form as
$$
\b \Psi=\diag((\b 1\tr \b \Pi)^{-1}) \b\Pi \diag(\b N)
$$
where $\b N$ is the vector of observed frequencies and $\b\Psi$ the matrix containing the reconstructed frequencies with the same arrangement as $\b \Pi$. Finally let $\bar{\b j}_t$ be the vector of indices that transforms $\b\Psi$ into the vector $\b\psi$ with entries in the lexicographic order associated with the DAG.

In the M-step we need to maximize the multinomial likelihood for the vector of reconstructed frequencies $\b\psi$. Because the joint distribution in a DAG can be factorized into the conditional distribution of each node $Z_i$ conditionally on its parents, the M-step may be split into a collection of smaller maximization problems. Though the theory for computing the corresponding estimates is more complex than fitting ordinary log-linear models due to the presence of more general link functions and, possibly, individual covariates, it is now well established within the field of marginal models: two efficient implementations suitable for our context are described in \cite{Eva-For}. To implement this methodology we construct, for each node, a vector of rearrangement indices $\b z_i$ which transforms $\b\psi$ into the marginal distribution of $(Z_i,pa_i)$ (a set of variables which may not be adjacent in the ordering of the DAG). Once the maximum likelihood for each node has been computed, we have a collection of vectors which contain the estimated conditional probability of each node given its parents; to reconstruct the full joint distribution $\hat{\b\pi}$, we use again a collection of vectors of rearrangement indices to transform the conditional distribution of each node properly expanded, into the original lexicographic order of the DAG.

In practice, the collection of rearrangement indices are set up at the beginning by a specific function. The whole collection of {\sc MatLab} functions that implement the EM algorithm on a general DAG will be made available as supplementary material. With some expertise, the models described in this paper could also be fitted with the LG-Syntax module described by \cite{VerMag:08}

To start the algorithm, an initial E-step is performed by assuming that the posterior probabilities $\pi_{\bl h \mid\bl j}(i)$ are uniform, except for a small random perturbation. In the initial M-step a one-step ahead logistic model is fitted and estimates are adjusted to smooth possibly large absolute values. In this way an initial estimate of the latent distribution is obtained.

The methodology described by \cite{BarFor:06}, section 3.3, is used to compute standard errors of the parameter estimates from the estimate of the expected information matrix. The idea is to collect all parameters into the vector $\b\beta$ and to compute the expected information matrix by the chain rule as follows
$$
\b F=\frac{\partial \b\theta\tr}{\partial\b\beta} \frac{\partial \b\gamma\tr}{\partial\b\theta}E\left( -\frac{\partial^2 L(\b\beta)}{\partial\b\gamma \partial\b\gamma\tr}\right) \frac{\partial \b\gamma}{\partial\b\theta\tr}\frac{\partial \b\theta}{\partial\b\beta\tr},
$$
where $\b\gamma$ is the vector of log-linear parameters for the saturated log-linear model of the observed distribution, $\b\theta$ is the vector of log-linear parameters for the latent distribution and $\b F$ is the expected information matrix. The implementation of this procedure for a general DAG model is a rather complex task which can be handled, again, by a specific routines which exploit the rearrangement indices mentioned above.
\section{Evaluation of causal effects}
\label{sec:causal}
Once a causal DAG has been estimated, seveal questions of interest can be formulated within the formal language developed by J. Pearl  \citep[see for example][Chapter 3]{PearlB} which we summarize briefly below. This formal language allows to evaluate causal effects by taking into proper account the causal relations described by the DAG and the role of unobserved confounders. However, while in a non-parametric context certain causal effects may not be estimable from the joint distribution of the observable variables, because our model is semi-parametric and identifiable, any causal effect of interest may be easily computed from the estimated latent distribution.

In a model defined by a system of structural equations, we can evaluate the causal effect of setting a subset of variables $X=(Z_i)_{i\in I}$ to $\b x$ on the probability that $Y=(Z_j)_{j\in J}$ equals $\b y$, with $J$ disjoint from $I$, by the so called "do operator": this is equivalent to determine the distribution that would arise if we could perform an idealized experiment where the variables in $\b X$ were randomized. The expression below is based on the fact that the joint distribution factorize into the product of the conditional distribution of each node but factors where the elements of $\b X$ are determined by their parents must be removed; the summation serves to marginalize relative to $\b Y$,
$$
 P(\b Y\mid do(\b X))=\sum_{i\not\in J} \prod_{i\not\in I} P(Z_i,\mid
pa_i).
$$
Once the intervention distribution has been computed, we may choose how to compare distributions of $Y$ for different values of $\b x$: the two most obvious alternatives are
differences or ratios of the relevant probabilities. Because in the application we deal with ordered categorical distributions, we compute the ratio of the corresponding survival probabilities.
\subsection{Direct effects}
In a complex DAG causal effects may travel through different pathways, thus $X$ may affect $Y$ directly or by affecting other variables which have a direct effect on $Y$. Consider, for instance, the model described in Figure \ref{simplifiedDAG} presented in section \ref{model}: $S^p$ (parents' education) affects $S^c$ (child education) directly, or by affecting $U^c$ (child latent endowment) or $I$ (family income) which, in turn, act on $S^c$. The effect of $U^p$ (parents' latent endowment) travels through many channels, but we would mainly be interested in its effect on $S^c$ while observed family backgrounds is held fixed, to capture the effect of natural inheritance, that is the path from $U^p$ to $S^c$ going through $U^c$.

Effects exerted through specific paths are called `direct effects'. In the literature different definitions of direct effects have been considered; the one used in our application is the `natural direct effect' which is defined as follows (see for example \citet{PearlB} Definition 4.5.1 or \citet{PearlNDE} section 6.1.3). Suppose we are interested in the causal effect of $\b X$ on $\b Y$ exerted through all paths except those going through a set of mediating variables $M=(Z_i)_{i\in K}$, with $K$ disjoint from $I,J$. First we computes the intervention distribution as if we were setting $X=\b x$ and $M=\b m$
$$
 P(Y\mid do(\b X),do(\b M))=\sum_{i\not\in J} \prod_{i\not\in I\cup K} P(Z_i,\mid pa_i),
$$
then we average this probability with weights provided by the
distribution of $M$ when $X$ is set to its reference category by
intervention.

Computation of direct effects requires the computation of several intervention distributions, a task that is similar to the one implemented within the M-step when we reconstruct the joint distribution. In practice, the basic ingredients are the DAG structure and, for each node, the estimated conditional distribution given its parents. Then, nodes are processed one at a time to compute the required intervention distribution. The software mentioned above has specific functions for this task.
\section{Application to Education transmission}
\label{sec:app}
The question of assessing the effect of raising the education of the parents by policy intervention on the education of their children is difficult because the answer depends on the extent to which the association between parents' and children's education is due to the transmission of unobservable endowments across generations.
\subsection{Background and Literature}
For simplicity, consider the very simple model in the four variables $S^p, U^p$ and $S^c, U^c$, which denote schooling and unobservable endowments respectively for parents and child, while $\epsilon^p, \:\epsilon^c$ are exogenous errors and assume that
\begin{eqnarray}
&U^c = g(S^p, U^p,\epsilon^p)\, \label{eq2} \\
&S^c = f(S^p, U^c,\epsilon^c). \label{eq1}
\end{eqnarray}
The first equation says that the child's endowment depends on parents' schooling and endowment while the second says that a child's education depends on her own endowment and on her parents' education.
Under this model the observed association between $S^p$
and $S^c$ is partly due to the effect of endowment on schooling
within each generation combined with the transmission effect from
$U^p$ to $U^c$. Thus the stronger the endowment transmission
effect the weaker the scope of education policy.
If we substitute from equation \eqref{eq2} into \eqref{eq1} we get the reduced form equation
\begin{equation}
S^c = f(S^p, U^p,\epsilon) \label{total}
\end{equation}
which requires controlling only for parents' endowment.
Three main approaches in this direction have been pursued. \citet{BehRos} study differences between subjects with twin mothers, having adjusted for assortative mating in order to control for differences between education of fathers; \citet{Plug} uses data on adoptees under the assumption that there should be no endowment transmission, although, as noted by \citet{survey}, association may be induced by selective placement of adoptees. \citet{BlaDevKje} analyze a dataset where differences in parent's education was exogenously induced by reforms in municipal schooling laws which may be treated as an instrumental variable. For a critical assessment see \cite{survey} who apply the three methods to a single data set and show that they produce conflicting results.

By fitting a much more complex version of \eqref{eq1}, Cameron and Heckman (1998) address the issue of how the family background affects the probability of transition from one grade of education to the next. Though their model resembles (\ref{eq1}) the heterogeneity is assumed independent from the observed covariates, so it could be interpreted as the component of $U^c$ which is not determined by family background.

The variable $U^p$, named {\em family endowment}, is essentially identified by the variables it affects, so it is meant to capture the family environment in which children grow up. It is in principle a cross classification of various characteristics of the family, but in practice it turns out to be naturally ordered in a scale of `quality'. The child's unobservable $U^c$ is identified mainly through cognitive and non-cognitive test scores, so it is not to be interpreted as strictly reflecting an individual intrinsic endowment; it is rather a mixture of this and other unobservables like motivation and acquired knowledge useful for schooling advancement.
\subsection{Data}
\label{data}
We use data from the National Child Development Survey (NCDS), produced by a UK cohort study targeting the population
born in the UK between the 3rd to the 9th of March 1958. Individuals
were surveyed at different stages of their life and information on
their schooling achievement, various tests results and family
background was collected. A complete description of the data is
available at \\
\texttt{http://www.esds.ac.uk/longitudinal/access/ncds}.

Some variables are inherently discrete (notably schooling level)
while those which were continuous, like income and test scores, were turned into discrete. Though a continuous variable contains more
information relative to a discrete approximation, there are two
reason why a model based on categorical variables may involve less
parametric restrictions than one based on the original continuous
measurements. When a continuous variable used as explanatory in a
regression model, non linearity can be achieved by assuming a given form, for instance polynomial; instead, once it has been transformed into a set of dummy variables corresponding to discrete categories, it can capture patterns of non linearity in a non parametric way. Models involving a continuous variable as response are usually based on the rather restrictive assumption of normality while, when used as categorical, the discrete distribution is assumed to be multinomial, that is completely unrestricted, at least in the first stage.

The original sample contains 18560 observations, but more than 80\% have at least a missing entry. Incompleteness is scattered across  many variables included in the survey. The subsample of complete data which we analyze amounts to almost 3000 subjects, 1471 males (sons) and 1330 females (daughters). The marginal distributions of the summary statistics for the most relevant variables in the complete-case sub-sample do not differ significantly from the same distributions in the whole sample, but we cannot really exclude selection bias. Our main dependent variable $S^c$ is the amount of education achieved by each individual, which takes four levels: no qualification, O-level, A-level and higher education.

Children are tested at the age of 7 and 11 for mathematics, reading
and non-cognitive skills, and again at 16 for math and reading, and
we use these test scores for identification of the unobservable
endowment. More specifically, we combine cognitive scores at 7 and 11 into two ordered variables: $M^e$, for Math, and $R^e$, for reading; Math and reading scores at 16 are coded into two additional variables
$M^l$ and $R^l$. Because {\em non-cognitive skills} (available at ages 7 and 11), were expressed by several measurements whose meaning was not entirely clear, they were summarized and then dichotomized at the median into the binary variable $N^c$.

Parents' schooling is defined as the age at which they left school (12 to 21 years); for each parent we extract a three level variable corresponding to significant educational steps: leaving up to 14 years of age; after 14 but not later than 16; after 16; these are called $S^m,\: S^f$ for mother and father respectively. As usual there are many missing data on family income; to alleviate the problem, since few mothers in the dataset have an income, we neglect mother's income (thus avoiding to drop data with missing mother's income) and concentrate on fathers'. We group their income in three categories into the ordered variable $I$.

The NCDS contains also information on parents' concern in their
children's education, as reported by teachers; this turns out to be
an important variable; it should measure the amount of effort/concern, and, perhaps, is related to the value that the family gives to the child's education. Parents' concern is originally classified into
as many as $5$ categories; we extract three binary parents'
variables, based on the age at which they were assessed $C^7,\:C^{11},\:C^{16}$.
\section{The Model}
\label{model}
The system of structural equations in the model are represented in the DAG of Fig 1 where, for simplicity, we have collapsed $S^m,\:S^f$ into the single node $S^p$ which has the same parents and descendants; in addition, there is an arrow from $S^m$ to $S^f$ to account for assortative mating.
\begin{figure}
\caption{The DAG used in the application}
\[
\xymatrix{ & & U^p \ar@{->}[lldd] \ar@{->}[dd] \ar@{->}[rd] \ar@{->}[ldd] \ar@{->}[rrd] \ar@{->}[rrrd]  \\
& & & C^7  \ar@{->}[dl] & C^{11}  \ar@{->}[dll] & C^{16}  \ar@{->}[dlll] & M^e \ar@{->}[r] & M^l \\
I \ar@{->}[rrd] & S^p \ar@{->}[l] \ar@{->}[r] \ar@{->}[dr] & U^c \ar@{->}[d] \ar@{->}[rrrru] \ar@{->}[rrrrru] \ar@{->}[rrrr] \ar@{->}[rd] \ar@{->}[rrd] & & & & N^c \\
& & S^c & R^e \ar@{->}[r] & R^l}
 \label{simplifiedDAG}
\]
\end{figure}
Additional features of the model are summarized in table \ref{Tmod} below where, for each variable in the DAG we give the number of categories and the type of logit ($g$ for global and $a$ for adjacent); whenever a node has two or more parents, their effects are assumed to be additive on the corresponding logit scale.
\begin{table}[htb]
\caption{Endogenous variables: number of categories and logits} 
\label{Tmod}
\centering 
\begin{tabular}{cllllllllllllll}  
\hline\noalign{\smallskip}
$Z_i$ & $U^p$ &$C^7$ &$C^{11}$ &$C^{16}$ &$S^m$ &$S^f$ & $I$ &$U^c$ & $M^e$ &$M^l$ &$R^e$ &$R^l$ &$N^c$ &$S^c$ \\
$c_i$ & 3& 2& 2& 2& 3& 3& 3& 3& 3& 3& 3& 3& 2& 4\\
Logit &a &g &g &g &g &g &g &a &g &g &g &g &a &g\\
\noalign{\smallskip}\hline
\end{tabular}
\end{table}
Because all observable variables in the system are naturally ordered, we use cumulative (or `global') logits; a cumulative logit could also be appropriate for $S^c$, the educational achievement of the child, however, a global link seemed to provide a substantially better fit. The levels of unobservable variables, instead, are assumed to correspond to unordered qualitative types, so we use adjacent logits. Separate models were fitted for daughters and sons to account for gender effects.
\begin{table}[ht]
\caption{Parameter estimates and standard errors(se) for sons and daughters}
\label{t:global_main}
\centering 
\begin{tabular}{lllccccc}  
\hline\noalign{\smallskip}
$Z_i$ & $pa_i$ & $h$ & \multicolumn{2}{c}{Daughters}  & & \multicolumn{2}{c}{Sons}  \\
\noalign{\smallskip}\hline
& & & coeff & se & & coeff & se  \\[.5ex]
\noalign{\smallskip}\hline
$C^{16}$ & $U^p$ & 1 &  -1.4133 &  0.3007 & & -1.3860 &   0.2721\\
$C^{16}$ & $U^p$ & 2 &  -2.7674 &  0.3951 & & -2.2284 &   0.2546\\	
$S^m$ & $U^p$ & 1 &  -2.5027 &   0.2776	& & -2.4058 &  0.2285	\\
$S^m$ & $U^p$ & 2 &  -0.7036 &  0.1750	& & -0.6301 &  0.1437	\\
$S^f$ & $U^p$ & 1 &  -2.9051 &  0.3717 & &  -5.7254 &  0.5329 \\
$S^f$ & $U^p$ & 2 &  -0.2087 &  0.2082 & &  -0.4647 &  0.1749 \\	
$U^c$ & $U^p$ & 1 &   1.2421 &  0.6426 & &   0.1621 &  0.5733\\
$U^c$ & $U^p$ & 2 &   2.8051 &  0.5843 & &   1.4549 &  0.4474\\
$U^c$ & $C^{16}$ & 1 &-0.5668 &  0.2332 & &  -0.8648 &  0.1737\\
$U^c$ & $S^m$ & 2 & -0.6575 &  0.2957 & &  -0.8067 &  0.2468\\
$U^c$ & $S^f$ & 2 & -0.0092 &  0.3903 & &  -0.6447 &  0.4554\\
$M^e$ & $U^c$ & 1 & -2.8296 &  0.1876& &   -2.8735 &  0.2047\\
$M^e$ & $U^c$ & 2& -2.6018 &  0.1278& &   -2.5633 &  0.1204\\			 $M^l$ & $U^c$ & 1 &-3.1387 &  0.2905& &   -3.0296 &  0.2333\\
$M^l$ & $U^c$ & 2 &-2.0644 &  0.2211& &   -1.5633 &  0.2147\\
$M^l$ & $M^e$ & 1 & 0.3476 &  0.1984& &    0.5046 &  0.1842\\
$S^c$ & $S^m$ & 1 & 0.1585 &  0.1491	& & -0.0599&   0.1422	\\
$S^c$ & $S^m$ & 2 & 0.3622 &  0.2009& &   -0.1078 &  0.2000\\
$S^c$ & $S^f$ & 1& 0.0818 &  0.1539& &    0.3573 &  0.1464\\
$S^c$ & $S^f$ & 2& 0.1118 &  0.2050& &    0.6453 &  0.2129\\
$S^c$ & $U^c$ & 1& -1.8843 &  0.1831& &   -3.0141 &  0.2135\\
$S^c$ & $U^c$ & 1& -2.8394 &  0.2107& &   -1.7979 &  0.2177\\
\noalign{\smallskip}\hline
\end{tabular}
\end{table}
\subsection{Main Estimation Results}
\label{coefficients}
\begin{table}[thb]
\caption{Some causal effects on $S^c$} \label{t:dirSc} \centering
\footnotesize
\begin{tabular}{lrrrr}
\hline\noalign{\smallskip}
 &  \multicolumn{2}{c}{Daughters} &
 \multicolumn{2}{c}{Sons}  \\
 & $S^c>0$ & $S^c>2$ & $S^c>0$ & $S^c>2$ \\
\noalign{\smallskip}\hline\noalign{\smallskip}
$U^p$ from 2 to 0 with & -& -& -& - \\
$I,S^c,C^7,C^{11},C^{16}$ fixed & 1.6869 & 3.7211 & 1.1920 & 1.5882 \\
$C^7$ from 0 to 1  & 1.0398 &    1.0733 &  1.0780   &  1.1263\\
$C^{11}$ from 0 to 1 & 1.1761   &  1.3554 & 1.2824  &  1.5182\\
$C^{16}$ from 0 to 1 & 1.1023   &  1.1943 & 1.2373  &  1.4215 \\
$S^m$ from 0 to 1 & 0.9383  & 0.9144 &
  0.9893    &  0.9699  \\
$S^m$ from 1 to 2 & 1.2005 &   1.5480 &
  1.1840  &  1.2741 \\
$S^m$ from 0 to 1 & 1.0593  &  1.1330 &
  1.1746  &  1.4384\\
$S^m$ from 1 to 2 &  1.0259  &  1.0748 &
  1.3195   & 1.9446\\
$I$ from 0 to 1 & 1.0275   &   1.0773 &
 1.0071   &  1.0178  \\
$I$ from 1 to 2 & 1.0546   & 1.1596 &  1.0720 &   1.1844  \\
\noalign{\smallskip}\hline
\end{tabular}
\normalsize
\end{table}
In Table \ref{t:global_main} we display some of the most relevant parameter estimates from different structural equations included in the model which we fitted to data on sons and daughters separately.
First of all note that all the coefficients for the effect of $U^p$ are negative and usually significant. This implies that, for instance, the probability that either parent has higher education decreases when their latent ability goes from 0 to 1 and from 1 to 2: this indicates that latent classes are ordered from best to worst. On the other hand, because the coefficients for the effect of $U^p$ on $U^c$ are positive, the latent ability of the child are probably ordered in the same direction. This result is in agrement with the fact that all the other coefficients in the $U^c$ equation are negative, indicating that increasing concern and higher education on the parents' side are positively associated with an improvement of the latent endowment of the child. The few displayed parameters from the equations for early and late score in math indicate that better endowed children get better score and that performances are correlated in time because, for instance, the coefficient relating $M^l$ to $M^e$ is positive. Finally, for $S^c$, the educational achievement of the child, the displayed estimates confirm that more endowed children have better chances of achieving higher education. The parameter estimates for the association with the education of the parents are less obvious to interpret: essentially we see that while the association with the education of the father is positive and significant for the son, the association for the daughter is close to 0 and smaller than the association with the education of the mother. These results may be interpreted as indicating a possible gender (or role) effect which may act either as pressure from the related parent or as an effort of emulation. A more specific interpretation of these results within the context of causal inference is described in the next section.
\subsection{Estimated causal effects}
As an application of the approach described in Section \ref{sec:causal}, certain causal effects on $S^c$ are displayed inTable \ref{t:dirSc} separately for sons and daughters. More precisely, the table contains the ratios
$$
\frac{P(S^c>s\mid do(x_1))}{P(S^c>s\mid do(x_0))}
$$
where $h=0,\:2)$ and $x_0, x_1$ denote the values of $X$ we want to compare. Because $S^p$ can affect $S^c$ directly or through $U^c$, we here report the direct effect only. Entries greater than 1 indicate positive effects of different strength while values close to 1 or even smaller may be interpreted as absence of any effect.

Though $U^p$ clearly cannot be manipulated, it is still useful to measure the effect of setting it to its highest/lowest value on $S^c$ when all family background variables are kept fixed. Note that the concern of the parents seems to have positive effects which are larger for daughters than for sons. The effect of changing mother's schooling from 0 to 1 seems to have no (or negative) effect while changing from 1 to 2 has a positive effect which is substantial mainly on daughters. On the other hand, changing father's education from 0 to 1 or from 1 to 2 has a positive effect which, however, is stronger on sons. Finally, the effect of income is positive, but, apparently rather weak.
\bibliographystyle{spbasic}      
\bibliography{TwoLat}   

\end{document}